\documentclass[conference]{IEEEtran}
\IEEEoverridecommandlockouts
\usepackage{amsmath,amssymb,amsfonts}
\usepackage{algorithmic}
\usepackage{graphicx}
\usepackage{textcomp}
\usepackage{xcolor}
\def\BibTeX{{\rm B\kern-.05em{\sc i\kern-.025em b}\kern-.08em
    T\kern-.1667em\lower.7ex\hbox{E}\kern-.125emX}}

\usepackage{graphicx} 

\usepackage[
    backend=biber,
    style=ieee,
    maxbibnames=20,
    doi=false,
    sorting=none
  ]{biblatex}
\title{Sealing the Deal: Effects of Fabrication Parameters on the Performance of Textile Pneumatic Haptic Actuators\\

\thanks{This work was supported in part by ******* grant *******.}
}

\author{\IEEEauthorblockN{Megan C. Coram}
\IEEEauthorblockA{\textit{Department of Mechanical Engineering} \\
\textit{Stanford University}\\
Stanford, USA \\
mccoram@stanford.edu}
\and
\IEEEauthorblockN{Allison M. Okamura}
\IEEEauthorblockA{\textit{Department of Mechanical Engineering} \\
\textit{Stanford University}\\
Stanford, USA \\
aokamura@stanford.edu}
\and
\IEEEauthorblockN{Cosima du Pasquier}
\IEEEauthorblockA{\textit{Department of Mechanical Engineering} \\
\textit{Stanford University}\\
Stanford, USA \\
cosimad@stanford.edu}
}

\begin{document}

\maketitle
\thispagestyle{empty}
\pagestyle{empty}

\begin{abstract}

Textile pneumatic actuators can provide useful wearable haptic feedback when embedded in gloves, armbands, and other smart garments. Here we investigate actuators fabricated from thermoplastic coated textiles. We measure the effects of fabrication parameters on the robustness and airtightness of small, round pneumatic pouch actuators made from heat-sealed thermoplastic polyurethane-coated nylon. We determine the optimal temperature, time, and pressure for heat-pressing of the textile to create strong bonds and identify the most effective glue to create an airtight seal at the inlet. Compared to elastomeric pneumatic actuators, these textile pneumatic actuators reduce the thickness of the actuator by 96.4\% and the mass by 57.2\%, increasing their wearability while maintaining a strong force output. We evaluated the force output of the actuators, along with their performance over time. In a blocked force test, the maximum force transmission of the pneumatic textile actuators was 36.1~N, which is 95.3\% of the peak force output of an elastomeric pneumatic actuator with the same diameter and pressure. Cyclical testing showed that the textile actuators had more stable behavior over time. These results provide best practices for fabrication and indicate the feasibility of textile pneumatic actuators for future wearable applications. 

\end{abstract}


\section{Introduction}
\label{Section I}

\begin{figure*}
\centering
\includegraphics[width=\textwidth]{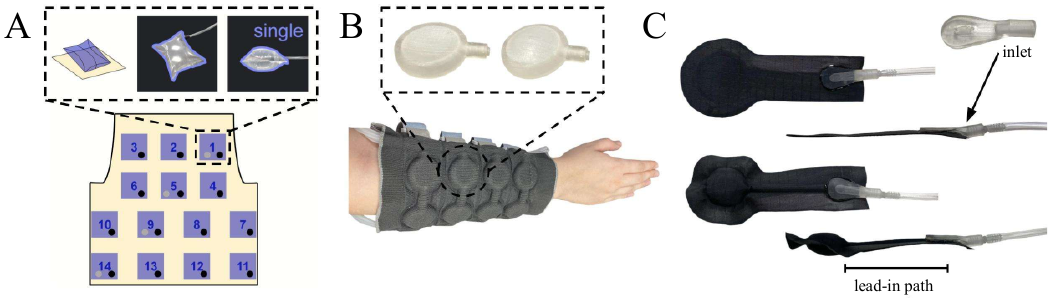}
\caption{A. Thermoplastic pouch pneumatic actuators in a vest \cite{Nunez2022AActuation}. B. 3D-printed elastomeric pneumatic actuators in a sleeve \cite{duPasquier2023ASleeve}. C. The textile pneumatic actuators tested in this paper, including an enlarged view of the 3D-printed inlet.}
\label{Fig. 1}
\end{figure*}

Pneumatic actuators are often preferred over other types of soft robotic actuation due to their high strength-to-weight ratio \cite{Jang2023DesignActuation}\cite{Niiyama2014PouchRobotics}. Textile and film-based actuators are slim and lightweight pouches, which can conform to the body and produce large force outputs \cite{ONeill2017ATesting}. A thin and flexible pouch inflates into a large and stiff volume; this transformation is particularly attractive for wearable haptic devices. 

Pneumatic wearable haptic devices have emerged as a promising alternative to vibration-based systems, offering a broader and more nuanced range of haptic feedback \cite{duPasquierHaptiknit:Haptics}. Using pneumatic pouches in contact with the skin, these devices mimic the feeling of human touch \cite{Talhan2024SoftTouch}, provide situational awareness \cite{Raitor2017WRAP:Guidance}\cite{Rognon2019SoftDrone}, and enhance immersion in virtual reality \cite{Delazio2018ForceExperiences}. Pneumatic actuators have been incorporated into bracelets \cite{Pohl2017Squeezeback:Notifications}\cite{Jumet2022ADevices}, sleeves \cite{Zhu2020PneuSleeve:Sleeve}\cite{Liu2021ThermoCaress:Stimulation}\cite{Choi2023DevelopmentActuator}\cite{Jumet2023FluidicallyTextiles}, fingertip bands \cite{VanBeek2024ValidationFeedback}\cite{Frediani2020TactileFingertip}, leg bands \cite{Endow2021Compressables:Interfaces}\cite{Fan2009PilotAmputee}, and vests \cite{Nunez2022AActuation}\cite{Delazio2018ForceExperiences}\cite{Jadhav2023ScalableActuators}\cite{Kang2023PneumaticSimulation} to provide a wide variety of cues and sensations, ranging from comforting human touch to collisions in virtual reality games. 

Many pneumatic actuators for haptic devices are made of thermoplastic films (polyethylene (PE), polyurethane (TPU), PVC, or similar) \cite{Niiyama2014PouchRobotics}\cite{Niiyama2015PouchDesign}\cite{Sanchez2019DevelopmentInteraction}\cite{Yamaoka2018AccordionFab:Sheets}\cite{Connolly2019Sew-freeRobots}\cite{Do2021Macro-MiniDisplays}\cite{Wu2019WearableArm}, cast elastomers \cite{Choi2023DevelopmentActuator}\cite{Endow2021Compressables:Interfaces}\cite{Kang2023PneumaticSimulation}, or coated textiles \cite{Jumet2023FluidicallyTextiles}\cite{Ou2016AeroMorphDesign}. Thermoplastic film tubes are particularly attractive because they are inexpensive and straightforward to adapt into rectangular pouches \cite{Nunez2022AActuation}\cite{Do2021Macro-MiniDisplays}. They can be embedded into wearable devices, such as the vest shown in Fig.~\ref{Fig. 1}A, thanks to their thin profile. Cast elastomeric pneumatic actuators are considerably more bulky, but can stretch to inflate farther than inextensible materials \cite{Choi2023DevelopmentActuator}\cite{Endow2021Compressables:Interfaces}. 3D-printed elastomeric pneumatic actuators can be fabricated with complex custom geometries suited to specific applications or fitting specific regions of the body. 

However, these thermoplastic and elastomeric actuators can cause discomfort and overheating when placed in direct skin contact for extended periods of time. Lack of breathability causes the user to sweat, which then causes the actuator to stick to the user's skin \cite{Winterhalter2012EffectsComfort}\cite{Zhong2006TextilesOverview.}. Coated textiles provide the same heat-sealing capabilities as thermoplastic films, but with a key advantage: unlike thermoplastic films, which are fully meltable, the textile substrate offers structural support, and only the coated side acts as an adhesive. Additionally, the uncoated side offers a soft fabric surface which is comfortable against the skin. Incorporating durable textiles like ripstop nylon enhances the longevity of the pouches while also enabling the actuators to be sewn directly into garments.
 
Many pneumatic devices require off-board air supplies in the form of heavy pumps or compressed air to achieve high bandwidth \cite{Jumet2023FluidicallyTextiles}. This limits their application in the home environment and as mobile devices. To make lightweight portable wearable devices, actuators should be low-volume and airtight to allow for operation by small, portable pneumatic sources, such as on-board pumps and CO2 cartridges. 
 
The effects of size and geometry on the performance of heat-sealed thermoplastic pouch actuators have been previously explored \cite{Jumet2023FluidicallyTextiles}\cite{Niiyama2015PouchDesign}\cite{Ou2016AeroMorphDesign}, but the effects of the parameters of the fabrication processes used to produce them have not yet been characterized. Methods of sealing including heat presses and heated rollers, both manual and computerized, have been proposed and contrasted, but variables such as temperature, pressure, and contact time have not yet been investigated. We propose that these settings affect the bond strength between the layers of the actuators, impacting their reliability and performance.

We present a parameterized fabrication process for modular textile pneumatic haptic actuators that are lightweight, low-profile, comfortable, and flexible, as shown in Fig.~\ref{Fig. 1}C. We focus on refining the fabrication process to ensure strong, airtight seals to produce durable actuators. We measure the performance of the actuators and compare them to the elastomeric 3D-printed actuators proposed by du Pasquier et al. \cite{duPasquierHaptiknit:Haptics} (Fig.~\ref{Fig. 1}B).

This paper is structured as follows: First, we present the fabrication method for the actuators and examine the results of varying parameters in the fabrication process. Then, we evaluate the performance of the textile pneumatic actuators and compare it against previous work. Finally, we contextualize the importance of this work in wearable haptic devices.

\section{Actuator Fabrication and \\ Parameter Evaluation}
\label{Section II}
\subsection{Fabrication}
\label{Section II.A}

\begin{figure*}
\centering
\includegraphics[width=\textwidth]{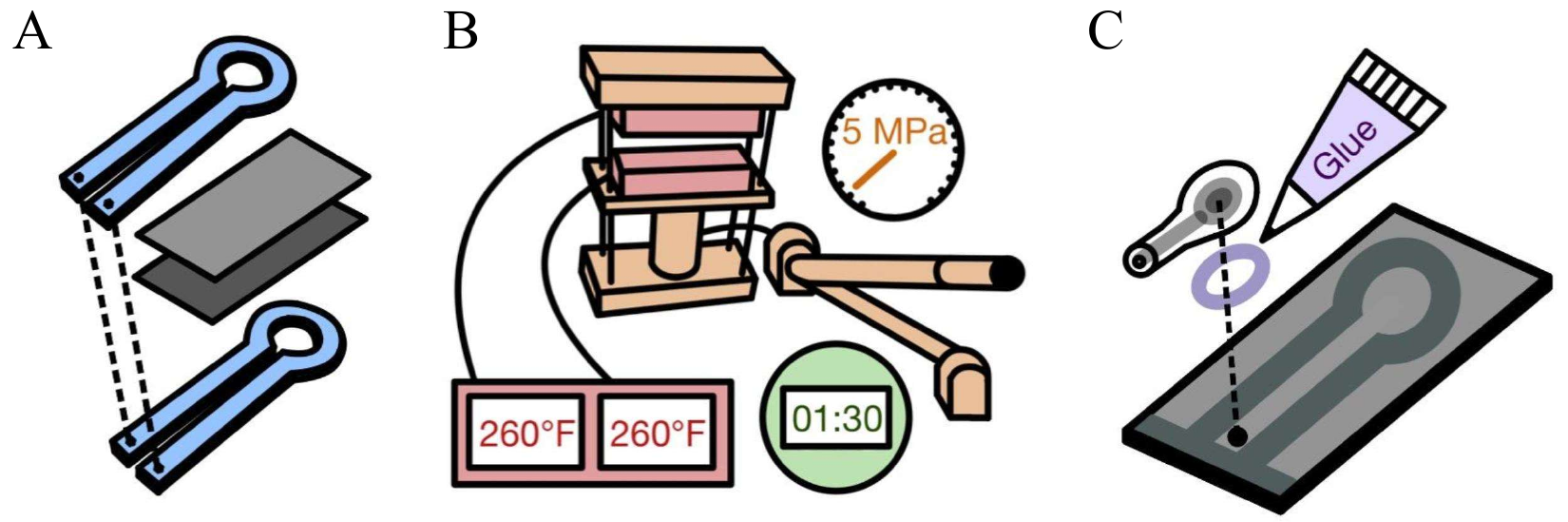}
\caption{The textile pneumatic actuator fabrication process. A. the textile sheets and the aluminum molds. B. The heat press and its settings. C. The 3D-printed inlet and bonding adhesive.}
\label{Fig. 2}
\end{figure*}

We fabricate our textile pneumatic actuators using TPU-coated 70-denier ripstop nylon (Quest Outfitters \#1056). Two pieces of nylon are placed in a heat press (DabPress Commercial Rosin Press) with the TPU-coated sides facing inwards. Above and below the fabric, we place 3/16” aluminum molds, as shown in Fig.~\ref{Fig. 2}A, which are in contact with the two heated plates of the press. These molds were waterjet cut to the desired geometry. Hexagonal pins ensure that the molds are aligned. Because the heat conductivity of the aluminum is higher than that of air, only the regions of the fabric that are in contact with the molds become sealed. The heat press is set to a target pressure, and the sample is kept in the press for a designated amount of time. After that time, the molds and sample are removed, the sample is set out flat to cool, and the molds are cooled with water to ensure a consistent starting temperature for every fabrication cycle. We discuss below in Section~\ref{Section II.C} how we selected the ideal temperature, pressure, and time.

Once cooled, elastomeric inlets must be glued to the actuators, as shown in Fig.~\ref{Fig. 2}C. The inlets are printed using Formlabs Flexible 80A resin to provide a slim and flexible connection to the air supply. The inlets are designed to provide a wide contact area for glue adhesion. A hole is pierced through one layer of the nylon actuator and is aligned with the hole in the inlet to allow for air flow. The types of glue and airtightness tests used to determine the best configuration are described below in Section~\ref{Section II.C}. Finally, we inflate each actuator to 200~kPa to ensure that the heat pressing has not sealed undesired regions of the actuator, and manually massage the actuator to ensure equal pressure distribution.

The round geometry of the actuator was selected to minimize stress concentrations around the edges of the air pocket. The choice of size was informed by previous literature, which found that voice coil \cite{Salvato2022Data-DrivenHumans} and elastomeric \cite{duPasquierHaptiknit:Haptics} actuators of this size could mimic the sensation of social touch on the dorsal forearm. The minimum spacing of the actuators, 25~mm center-to-center, is larger than the 2-point discrimination threshold of the lower dorsal arm \cite{Nolan1982Two-pointWomen}, the back \cite{Catley2013AssessingFoot}, and the legs \cite{Ercalk2021Two-pointIndividuals}. Thus, for haptic applications in these areas, there is no need for actuators to be smaller. Smaller actuators would also have a reduced indentation depth, resulting in less haptic feedback. Larger actuators might provide stronger haptic cues, but would require a stronger pump or take more time to inflate, resulting in a heavier or slower device. They would also have a larger contact area with the skin, which might reduce cue specificity. In addition to the main round pouch, the actuator geometry features a long, thin lead-in path that allows the inlet to be positioned separately from the main volume of the actuator.

\subsection{Parameter Evaluation}
\label{Section II.B}

The fabrication parameters that we investigate are the temperature and pressure of the heat press, the duration the sample was pressed, and the type of glue used to attach the inlets. Preliminary testing indicated that heat press temperatures exceeding 200~°F are required to fully bond the TPU-coated nylon, while temperatures above 300~°F seal areas outside of the intended contact area. To quantify the relationship between temperature and bond strength, we test temperatures between 220~°F and 280~°F at 20~°F intervals for a total of 4 temperatures. Previous literature \cite{Jumet2022ADevices} indicates that low pressures are required to bond the textiles, so the smallest intervals on our heat press are used. We test samples at three pressures: 2.5, 5, and 7.5~MPa. In previous literature \cite{Jumet2022ADevices}\cite{Ou2016AeroMorphDesign}, pressing durations of 30~s or less were used. To determine whether longer pressing durations would produce a stronger bond, we tested from 30 to 120~s at 30~s intervals for a total of 4 pressing durations. To reduce the number of combinations to test, we designate 260~°F, 5~MPa, and 90~s as nominal values for our testing ranges and vary one parameter at a time.

We determined the quality of the bond with respect to pressure, temperature, and pressing duration with peel tests. We ran our tests according to the ASTM D1876 T-Peel Test \cite{2023D1876Test}. Rectangular samples (n=6) of width 25~mm and length 120~mm (25~mm unbonded and 95~mm heat bonded) were produced for each combination of temperature, pressure, and time (a total of 9 conditions and 54 samples). These samples were created using solid rectangular aluminum molds. Peel tests were performed on an Instron 68TM-50 with a 5~kN 2580 series static load cell and screw side action tensile grips. The results are reported as force-displacement curves. We align the data at a 5~N pretension value to reduce the effects of slack in the clamping setup. To assess the quality of adhesion, we measure the resistance-to-peel strength (or peel strength) of the samples \cite{Bartlett2023PeelReview}. We calculate the peel strength as the average slope of the force-displacement line during its initial rise. We define a further metric to describe the point at which the samples begin to fail (analogous to the proportional limit in a stress-strain curve): the point at which the gradient falls below half of the calculated peel strength. We refer to the displacement at this point as the critical displacement.

We tested six different types of adhesives to mount the inlets: (1) \textit{Hardman Double/Bubble Blue General Purpose 2-Part Epoxy}, (2) \textit{Gear Aid Seam Grip +WP Waterproof Sealant and Adhesive} (a urethane-based adhesive), (3) \textit{Loctite Vinyl, Fabric, and Plastic Flexible Adhesive} (a polyurethane-based adhesive), (4) \textit{3M High Performance Acrylic Adhesive 200MP Tape} (a double-sided tape), (5) \textit{Gorilla Super Glue Gel} (a cyanoacrylate-based adhesive), and (6) \textit{Loctite Instant Mix 5-Minute Epoxy}. Using the assembly process described above, we attached inlets using each glue for n=4 samples and performed airtightness tests to evaluate the quality of the bond. For this, we inflated each actuator to 150~kPa, shut off the air supply, and monitored the pressure drop over time. To evaluate the variability of the samples, we calculated the average standard deviation in the pressure readings over the first 8 seconds while the actuators were actively deflating.

\begin{figure*}
\includegraphics[width=\textwidth]{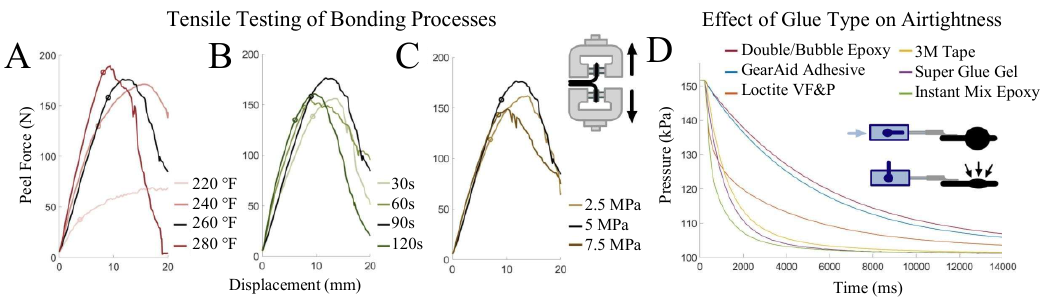}
\caption{Effects of varying the fabrication A. temperature, B. time, C. pressure, and D. adhesive on the peel strength and airtightness, with the nominal values of 260~°F, 90s, and 5MPa shown in black and the critical displacement points circled.}
\label{Fig. 3}
\end{figure*}

\begin{table*}
    \centering
    \caption{Metrics of samples created with different fabrication parameters}
    \label{Table 1}
    \renewcommand{\arraystretch}{1.2}
    \begin{tabular}{c|ccccccccc}
        Condition & Nominal$^{\mathrm{a}}$ & 220~°F & 240~°F & 280~°F & 30~s & 60~s & 120~s & 2.5~MPa & 7.5~MPa \\
    \hline
        Standard Deviation (N) & 14.89 & 16.84 & 8.28 & 37.41 & 19.28 & 18.48 & 28.97 & 9.43 & 11.02 \\
         Peak Force (N) & 176.08 & 68.83 & 171.27 & 189.22 & 155.79 & 154.97 & 160.45 & 161.75 & 148.72 \\
         Peel Strength (N/mm) & 17.82 & 7.49 & 19.24 & 24.49 & 15.91 & 20.29 & 22.60 & 17.19 & 18.74 \\
         Critical Displacement (mm) & 9.06 & 3.81 & 7.20 & 8.13 & 9.31 & 7.87 & 6.09 & 7.03 & 8.47\\
    \hline
    \multicolumn{4}{l}{$^{\mathrm{a}}$Nominal settings are 260~°F, 90s, and 5MPa.}
    \end{tabular}
\end{table*}

\begin{table*}
    \centering
    \caption{Standard deviation and cure time of adhesives}
    \label{Table 2}
    \renewcommand{\arraystretch}{1.2}
    \begin{tabular}{c|cccccc}
        Adhesive Type & Double/Bubble & Gear Aid & Loctite VF\&P & 3M Tape & Super Glue Gel & Instant Mix \\
    \hline
        Standard Deviation (kPa) & 0.33 & 1.45 & 13.21 & 1.89 & 2.20 & 3.54\\
         Cure Time (hours) & 24 & 12 & 24 & 0 & 24 & 24 \\
    \end{tabular}
\end{table*}

\subsection{Results and Discussion}
\label{Section II.C}

Varying the temperature to which the samples were heated showed that bonding temperature affects both the bond strength and the material durability. At 220~°F, the adhesive bond is much weaker than at higher temperatures, as shown by its low peel strength and low peak force in Fig.~\ref{Fig. 3}A and Table~\ref{Table 1}. This indicates that the 220~°F TPU does not get fully fused. At 240~°F, the bond provides a high peel strength and the samples fail at an 89\% higher displacement than those created at 220~°F because the adhesion is more complete. At 260~°F, the samples display a 7.4\% lower peel strength than those created at 240~°F, but have a 25.8\% higher critical displacement and reach a 2.8\% higher maximum force, as shown in Table~\ref{Table 1}. At 280~°F, the samples display the highest peel strength and highest maximum force. However, the samples sealed at 280~°F have a lower critical displacement than the samples sealed at 260~°F, indicating that the higher temperature causes the material to become more brittle. We choose to use 260~°F because it provides acceptable peel strength and reaches a high peak force without significantly weakening the material. 

The pressing duration for which the samples are heated has a similar effect on the bond strength as temperature. The samples pressed for 30~s display low bond stiffness, as shown by their low peel strength in Table~\ref{Table 1}. The samples pressed for 60~s have a 27.5\% higher peel strength than those pressed for 30~s, but experience failure at a 15.5\% lower critical displacement. The samples pressed for 90~s, however, display a peel strength in between those of the 30-second and 60-second samples, and achieve a 13\% higher maximum force. The 120-second samples have the highest stiffness but the lowest critical displacement amongst the time variations. We choose to use a pressing duration of 90~s to fabricate our actuators because of the superior strength it provided; its maximum force was 9.7\% higher than the next-highest maximum force among the time variations.

The heat press pressure also affects the bond strength. As shown in Fig.~\ref{Fig. 3}C and Table~\ref{Table 1}, all three pressure settings provide almost identical peel strengths. However, samples pressed at 2.5~MPa have a low critical displacement, and samples pressed at 7.5~MPa have a low peak force. Thus, 5~MPa is selected as the best pressure for fabrication based on its large critical displacement and high peak force.

As shown in Fig.~\ref{Fig. 3}D, the type of glue used in mounting the inlets impacts their airtightness. \textit{Hardman Double/Bubble Blue General Purpose 2-Part Epoxy} created the most airtight seal, as evidenced by the slow dropoff in air pressure of the corresponding curve in Fig.~\ref{Fig. 3}D. It also has the lowest variability between trials, as shown in Table~\ref{Table 2}, indicating high consistency between actuators. However, this epoxy comes in single-use packets which cannot be resealed, so it is inefficient for creating small batches of actuators, which only require a small amount of glue. In addition, the \textit{Double/Bubble} epoxy can peel off of the inlets when subjected to bending, which means it is not suitable for applications in which actuators could experience bending. 

\textit{Gear Aid Seam Grip +WP Waterproof Sealant and Adhesive} provides a similar level of airtightness as the \textit{Double/Bubble} adhesive, as shown by the closely-spaced lines in Fig.~\ref{Fig. 3}D. It exhibits the second-lowest variability amongst the glue types, as shown in Table~\ref{Table 2}, but its variability is 4.39 times that of the \textit{Double/Bubble} epoxy. This adhesive comes in a resealable tube, which allows for smaller batches to be produced without waste. The \textit{Gear Aid} adhesive does not peel off of the fabric or the inlets when bent, so it is better suited to applications where the actuators may be bent when worn. It also offers the lowest curing time amongst the liquid glues, as shown in Table~\ref{Table 2}.

\textit{Loctite Vinyl, Fabric, and Plastic Flexible Adhesive} displays the largest variability by a factor of 3.73 amongst the glue types, as shown in Table~\ref{Table 2}. Some of the samples show airtightness similar to the \textit{Double/Bubble} and \textit{Gear Aid} samples, while others drop rapidly in pressure. \textit{3M High Performance Acrylic Adhesive 200MP Tape} offers the lowest drying time and third-lowest variability, but does not create an airtight seal, resulting in a sharply-sloped line in Fig.~\ref{Fig. 3}D. \textit{Gorilla Super Glue Gel} and \textit{Loctite Instant Mix 5-Minute Epoxy} also produce sharply-sloped lines because they are not airtight. The \textit{Super Glue} appears to shrink and crack while drying, while the \textit{Instant Mix Epoxy} becomes rigid when it dries. These behaviors make it easy for the inlets to peel off, which likely causes the high variability seen in Table~\ref{Table 2}.

We choose to use the \textit{Gear Aid} adhesive to fabricate the textile pneumatic actuators because of its airtightness, its low variability, its economical packaging, and its smoothness and flexibility when dried.

Fig.~\ref{Fig. 3}D shows that all actuators experience a significant pressure drop-off over time. This does not necessarily indicate that bonding of the actuator or the inlet is not airtight. The fabric itself allows for some diffusion and the pressure control system also has leaks; when we replace the actuator with a closed seal, the control system continues to register a pressure drop-off.

\section{Actuator Performance}
\label{Section III}

\begin{figure*}[ht!]
\centering
\includegraphics[width=\textwidth]{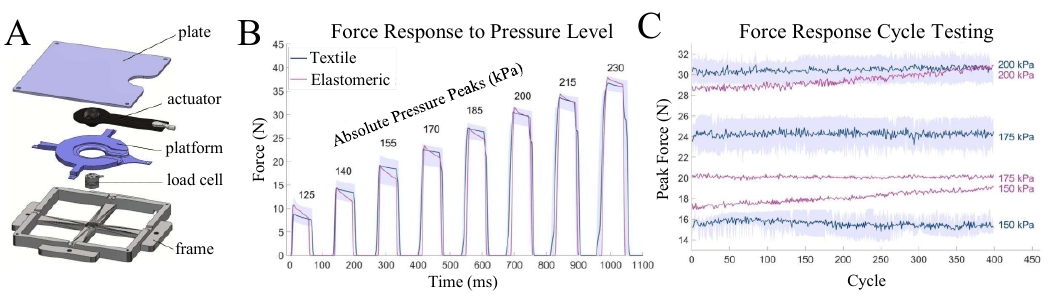}
\caption{A. Force testing setup, B. measured force at pressure peaks, and C. cyclical force testing results.}
\label{Fig. 4}
\end{figure*}

\subsection{Methods}
\label{Section III.A}

Using the fabrication parameters found above, we fabricated a set of textile pneumatic actuators. We assess the performance of these actuators in force transmission and in cyclical testing. We choose these tests to evaluate the ability to produce normal force, to compare to similar pneumatic actuators \cite{duPasquierHaptiknit:Haptics}, and to ensure that the chosen fabrication method is robust enough to be used for wearable haptic devices, in which each actuator will undergo thousands of cycles.
The force characterization tests (n=5) are conducted following a procedure similar to that outlined in \cite{duPasquierHaptiknit:Haptics} and adapted for textile pneumatic actuators. As shown in Fig.~\ref{Fig. 4}A, the actuator is placed on a 3D-printed platform with a hole in it for the actuator to expand through. When inflated, the actuator presses against a force sensor (ATI Nano17 6-axis Force/Torque Sensor) below. Above the actuator, an acrylic plate prevents the actuator from expanding away from the force sensor. The platform, force sensor, and acrylic plate are secured to a 3D printed frame. The actuator is inflated by a pneumatic controller similar to the one proposed by du Pasquier et al. \cite{duPasquierHaptiknit:Haptics}. During the test, the actuator is inflated to a series of absolute pressures from 125 to 230~kPa at 15~kPa intervals (a total of 8 peaks) and vacuumed to atmospheric pressure in between each pressure. The force is recorded at 100~Hz by the force sensor.

Cyclical testing (n=4) is performed with the same frame and load cell setup. Actuators are inflated to three target pressures, 150, 175, and 200~kPa, for 400 cycles. During each cycle, the actuator is inflated to the target pressure, held for one second, vacuumed to atmospheric pressure, and held for one second. Again, the force sensor records the output force at a frequency of 100~Hz. For each cycle, one peak force value is extracted from the data.

\subsection{Results}
\label{Section III.B}

In the characterization tests, our textile pneumatic actuators (represented in blue in Fig.~\ref{Fig. 4}A) displayed similar force transmission performance compared to 3D-printed elastomeric pneumatic actuators (represented in magenta in Fig.~\ref{Fig. 4}A). As shown by the shaded regions illustrating variability across the trials, the force output varies by a few Newtons from one actuator to another, but the average stayed close to that of the elastomeric pneumatic actuators. With the current pump, the pneumatic controller can provide up to 230~kPa (about 130~kPa above atmospheric pressure), which translates to over 36~N of force transmission. The blocked-force acrylic plate setup represents the actuator's optimal performance by directing all actuator deformation along the axis of the force sensor.

The cyclical tests indicate that our textile pneumatic actuators are robust and do not exhibit signs of degradation over time. The actuators do not stretch or peel; their shape and volume stay consistent. For a given target pressure level, the force peaks stay consistent within a range of 3.6~N across cycles of a single actuator and a range of 5.15~N across all samples tested at a given pressure, as shown by the shaded regions illustrating variability in Fig.~\ref{Fig. 4}C. The variability in force output during cyclical testing can be attributed to the reaction time of the controller or to the sampling rate of the force sensor.

Compared to the 3D-printed elastomeric pneumatic actuators from \cite{duPasquierHaptiknit:Haptics}, these textile pneumatic actuators have higher consistency in the cycling tests. At 150 and 200~kPa, the output force of the elastomeric pneumatic actuators drifted by a few Newtons over time. This is likely caused by the elastomeric actuators stretching and deforming with repeated use. 

\section{Discussion and Conclusions}
\label{Section IV}

We investigated the effects of several fabrication parameters on the process of creating heat-sealed textile pneumatic actuators: temperature, pressure, and pressing duration. We conclude that the best parameters for fabricating the actuators are to heat press the TPU-coated nylon at 260~°F for 90~s under 5~MPa of pressure. These parameters are specific to our setup. Different results may be obtained by varying factors such as the type or thickness of the textile or by using a heat press with a single heated plate instead of two. Independent of exact values, we show that temperature, time, and pressure can have significant effects on the peel strength, peak force, and critical displacement of heat-sealed textiles, and outline a benchmarking method to evaluate the best fabrication parameters for any given setup. We also find that \textit{Gear Aid} adhesive creates the best bond to our materials, and we can conclude more broadly that key features in the choice of an adhesive for textile pneumatic actuators are the following: repeatability, packaging, and post-cure adhesion, smoothness, and pliability.

In testing, our actuators produce up to 36.1~N at 230~kPa absolute pressure when placed in a blocked force testing setup, which is 95.3\% of the force produced by elastomeric pneumatic actuators of a comparable size in a similar setup \cite{duPasquierHaptiknit:Haptics}. In contrast, textile pneumatic actuators achieve a 96.4\% reduction in thickness when uninflated, decreasing the profile from 8.3~mm to 0.3~mm. The textile pneumatic actuators also achieve a 57.2\% reduction in mass, from 4.73~g to 2.03~g.
This allows our textile pneumatic actuators to be incorporated into lower-profile wearable devices without significantly compromising performance. Additionally, the textile actuators demonstrated more consistent performance over time during cycling tests, suggesting greater reliability for long-term use.

Jumet et al.\ pioneered similar textile pneumatic actuators in bracelets \cite{Jumet2022ADevices} and a sleeve \cite{Jumet2023FluidicallyTextiles}. For these devices, a series of actuators were produced by heat-pressing two sheets of TPU-coated nylon taffeta fabric, between which a laser-cut paper layer prevents bonding in the inflation areas. This paper layer is not removable, but it allows for more convenient customization in actuator size and shape, as well as patch repairs in the case of punctured actuators. Like the actuators we tested, these actuators are compliant, comfortable, and low-profile. Their 25~mm square cells offer a 625~mm$^2$ contact area, while our 25~mm round actuators offer a 491~mm$^2$ contact area. At 100~kPa (200~kPa absolute pressure), their actuators produce up to 9.8~N of output force in a test setup mimicking the stiffness of the human wrist. In this setup, some of the force is absorbed deforming the synthetic skin, and the actuator may expand away from the force sensor, resulting in a lower measured force than a blocked-force test. In our blocked-force setup, we measured up to 30.6~N at the same pressure, but without losses. Assuming skin deformation and outward expansion reduce the force output by no more than 68\%, our force measurements indicate that our actuator geometry is more effective at converting air pressure into force. These results also highlight the value of distributed-stiffness textiles, as described in \cite{duPasquierHaptiknit:Haptics}, for controlling the deformation of embedded pneumatic actuators. Layering stiff textiles over the actuator mimics a blocked-force setup by directing actuator deformation in a single direction.

\begin{figure}[h]
\includegraphics[width=0.5\textwidth]{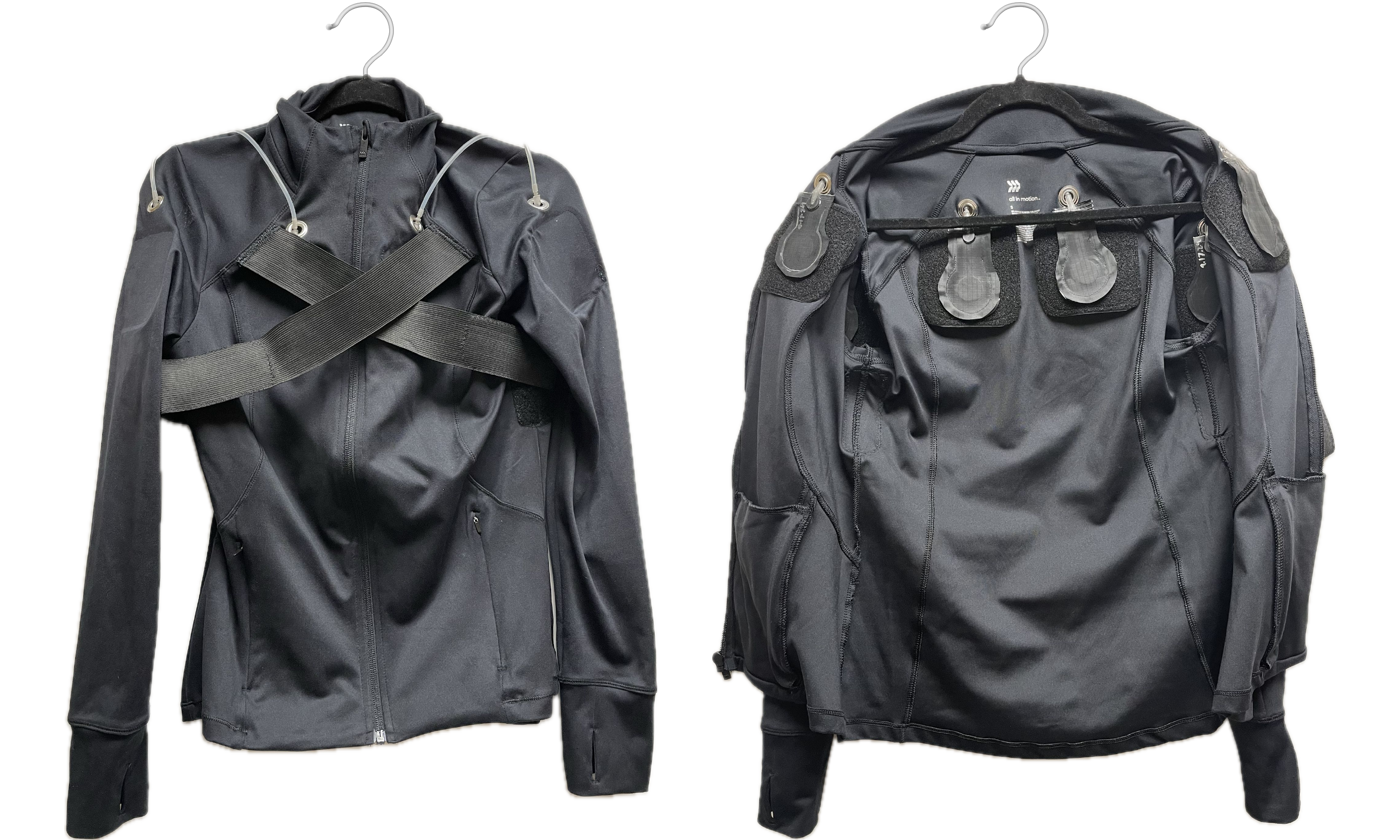}
\caption{Modular textile actuators integrated into a portable vest for haptic feedback on the torso.}
\label{Fig. 5}
\end{figure}

Due to to their low profile and rapid and repeatable fabrication method, textile pneumatic actuators can be integrated into a variety of wearable devices, such as bracelets, sleeves, collars, pants, or vests, as illustrated in Fig.~\ref{Fig. 5}. Their form factor allows for convenient and discrete integration into garments, increasing their wearability and user appeal. The soft textile surface of the actuators allows for comfortable nonadhesive direct-to-skin contact. Their compact inflation volume allows for operation with a portable controller equipped with an onboard pump and vacuum, thereby enhancing the potential for untethered wearability. Potential applications for pneumatic haptic garments with include improving immersion in virtual reality \cite{Delazio2018ForceExperiences}\cite{Jadhav2023ScalableActuators}\cite{Kang2023PneumaticSimulation}, replicating social touch in digital environments \cite{Talhan2024SoftTouch}\cite{Keng2008HuggySystem}, assisting navigation \cite{Raitor2017WRAP:Guidance}\cite{Jumet2023FluidicallyTextiles}, or providing guidance and feedback for posture or breathing \cite{Wu2010Torso-basedDisorders}\cite{Rotella2012HAPIInterface}.
 





\section*{Acknowledgments}

The authors thank Prof. [anonymous] for use of the Instron machine, Prof. [anonymous] for figure reuse, and [anonymous] for preliminary work and design of the testing apparatus.


\printbibliography

@inproceedings{duPasquier2023ASleeve,
    title = {{A Knit Haptic Pneumatic Sleeve}},
    year = {2023},
    booktitle = {IEEE World Haptics Conference (Demonstration)},
    author = {du Pasquier, Cosima and Scholl, Ian and Tessmer, Lavender and Tibbits, Skylar and Okamura, Allison},
    note = {https://youtu.be/s4SANSe34eY}
}

@article{Nunez2022AActuation,
    title = {{A Large-Area Wearable Soft Haptic Device Using Stacked Pneumatic Pouch Actuation}},
    year = {2022},
    journal = {IEEE/RSJ International Conference on Intelligent Robots and Systems},
    author = {Nunez, Cara M. and Do, Brian H. and Low, Andrew K. and Blumenschein, Laura H. and Yamane, Katsu and Okamura, Allison M.},
    pages = {591--598},
    volume = {},
    publisher = {Institute of Electrical and Electronics Engineers Inc.},
    doi = {10.1109/IROS47612.2022.9981919}
}

@article{ONeill2017ATesting,
    title = {{A soft wearable robot for the shoulder: Design, characterization, and preliminary testing}},
    year = {2017},
    journal = {IEEE International Conference on Rehabilitation Robotics},
    author = {O'Neill, Ciarán T. and Phipps, Nathan S. and Cappello, Leonardo and Paganoni, Sabrina and Walsh, Conor J.},
    pages = {1672--1678},
    publisher = {IEEE Computer Society},
    doi = {10.1109/ICORR.2017.8009488}
}

@article{Jumet2022ADevices,
    title = {{A Textile-Based Approach to Wearable Haptic Devices}},
    year = {2022},
    journal = {IEEE International Conference on Soft Robotics},
    author = {Jumet, Barclay and Zook, Zane A. and Xu, Doris and Fino, Nathaniel and Rajappan, Anoop and Schara, Mark W. and Berning, Jeffrey and Escobar, Nicolas and O'Malley, Marcia K. and Preston, Daniel J.},
    pages = {741--746},
    publisher = {Institute of Electrical and Electronics Engineers Inc.},
    doi = {10.1109/ROBOSOFT54090.2022.9762149}
}

@article{Yamaoka2018AccordionFab:Sheets,
    title = {{AccordionFab: Fabricating Inflatable 3D Objects by Laser Cutting and Welding Multi-Layered Sheets}},
    year = {2018},
    journal = {Symposium on User Interface Software and Technology},
    author = {Yamaoka, Junichi and Nozawa, Kazunori and Asada, Shion and Niiyama, Ryuma and Kawahara, Yoshihiro and Kakehi, Yasuaki},
    pages = {160--162},
    doi = {10.1145/3266037.3271636}
}

@article{Ou2016AeroMorphDesign,
    title = {{aeroMorph - Heat-sealing Inflatable Shape-change Materials for Interaction Design}},
    year = {2016},
    journal = {Symposium on User Interface Software and Technology},
    author = {Ou, Jifei and Skouras, Mélina and Vlavianos, Nikolaos and Heibeck, Felix and Cheng, Chin-Yi and Peters, Jannik and Ishii, Hiroshi},
    pages = {121--132},
    publisher = {ACM},
    doi = {10.1145/2984511.2984520}
}

@article{Catley2013AssessingFoot,
    title = {{Assessing tactile acuity in rheumatology and musculoskeletal medicine - how reliable are two-point discrimination tests at the neck, hand, back and foot?}},
    year = {2013},
    journal = {Rheumatology},
    author = {Catley, Mark J. and Tabor, Abby and Wand, Benedict M. and Moseley, G. Lorimer},
    number = {},
    pages = {1454--1461},
    volume = {52},
    publisher = {Rheumatology (Oxford)},
    doi = {10.1093/RHEUMATOLOGY/KET140}
}

@article{Endow2021Compressables:Interfaces,
    title = {{Compressables: A Haptic Prototyping Toolkit for Wearable Compression-based Interfaces}},
    year = {2021},
    journal = {ACM Designing Interactive Systems Conference},
    author = {Endow, Shreyosi and Morad, Hedieh and Srivastava, Anvay and Noya, Esau G. and Torres, Cesar},
    pages = {1101--1114},
    doi = {10.1145/3461778.3462057}
}

@article{2023D1876Test,
    title = {{ASTM D1876-08 Standard Test Method for Peel Resistance of Adhesives (T-Peel Test)}},
    year = {2023},
    journal = {American Society for Testing and Materials (ASTM) International},
    note = {DOI: 10.1520/D1876-08R23}
}

@article{Salvato2022Data-DrivenHumans,
    title = {{Data-Driven Sparse Skin Stimulation Can Convey Social Touch Information to Humans}},
    year = {2022},
    journal = {IEEE Transactions on Haptics},
    author = {Salvato, Mike and Williams, Sophia R. and Nunez, Cara M. and Zhu, Xin and Israr, Ali and Lau, Frances and Klumb, Keith and Abnousi, Freddy and Okamura, Allison M. and Culbertson, Heather},
    number = {2},
    pages = {392--404},
    volume = {15},
    publisher = {Institute of Electrical and Electronics Engineers Inc.},
    doi = {10.1109/TOH.2021.3129067}
}

@article{Jang2023DesignActuation,
    title = {{Design of Gusseted Pouch Motors for Improved Soft Pneumatic Actuation}},
    year = {2023},
    journal = {IEEE/ASME Transactions on Mechatronics},
    author = {Jang, Jae Hyuck and Jamil, Babar and Moon, Youngjin and Coutinho, Altair and Park, Gijun and Rodrigue, Hugo},
    number = {6},
    pages = {3053--3063},
    volume = {28},
    publisher = {Institute of Electrical and Electronics Engineers Inc.},
    doi = {10.1109/TMECH.2023.3244347}
}

@article{Sanchez2019DevelopmentInteraction,
    title = {{Development of airbag fabrication machine and process for physical human machine interaction}},
    year = {2019},
    journal = {IEEE International Conference on Soft Robotics},
    author = {Sanchez, Eric Sebastian and Zamora, Daniel Campos and Kim, Joohyung},
    pages = {502--508},
    publisher = {Institute of Electrical and Electronics Engineers Inc.},
    doi = {10.1109/ROBOSOFT.2019.8722726}
}

@article{Choi2023DevelopmentActuator,
    title = {{Development of human-touch smart armband for tele-haptic communication using a fabric-based soft pneumatic actuator}},
    year = {2023},
    journal = {Fashion and Textiles},
    author = {Choi, Hanna and Yoo, Shinjung},
    number = {},
    pages = {1--14},
    volume = {10},
    publisher = {Springer},
    doi = {10.1186/S40691-023-00350-Y}
}

@article{Winterhalter2012EffectsComfort,
    title = {{Effects of Overgarment Moisture Vapor Transmission Rate on Human Thermal Comfort}},
    year = {2012},
    journal = {ASTM Special Technical Publication},
    author = {Winterhalter, Carole and Truong, Quoc and Endrusick, Thomas and Cardello, Armand and Lesher, Larry},
    pages = {129--157},
    volume = {1544},
    publisher = {ASTM International},
    doi = {10.1520/STP104085}
}

@article{Jumet2023FluidicallyTextiles,
    title = {{Fluidically programmed wearable haptic textiles}},
    year = {2023},
    journal = {Device},
    author = {Jumet, Barclay and Zook, Zane A and Yousaf, Anas and Rajappan, Anoop and Xu, Doris and Yap, Te Faye and Fino, Nathaniel and Liu, Zhen and O'Malley, Marcia K and Preston, Daniel J},
    pages = {100059},
    doi = {10.1016/j.device.2023.100059}
}

@article{Delazio2018ForceExperiences,
    title = {{Force Jacket: Pneumatically-Actuated Jacket for Embodied Haptic Experiences}},
    year = {2018},
    journal = {ACM Conference on Human Factors in Computing Systems (CHI)},
    author = {Delazio, Alexandra and Nakagaki, Ken and Klatzky, Roberta L and Hudson, Scott E and Lehman, Jill Fain and Sample, Alanson P},
    pages = {320},
    volume = {},
    doi = {10.1145/3173574.3173894}
}

@article{Rotella2012HAPIInterface,
    title = {{HAPI bands: A haptic augmented posture interface}},
    year = {2012},
    journal = {IEEE Haptics Symposium},
    author = {Rotella, Michele F. and Guerin, Kelleher and He, Xingchi and Okamura, Allison M.},
    pages = {163--170},
    doi = {10.1109/HAPTIC.2012.6183785}
}

@article{duPasquierHaptiknit:Haptics,
    title = {{Haptiknit: Distributed Stiffness Knitting for Wearable Haptics}},
    journal = {Science Robotics},
    author = {du Pasquier, Cosima and Tessmer, Lavender and Scholl, Ian and Tilton, Liana and Chen, Tian and Tibbits, Skylar and Okamura, Allison},
    volume = {9},
    number = {97},
    year = {2024},
    doi = {10.1126/scirobotics.ado3887}
}

@article{Keng2008HuggySystem,
    title = {{Huggy Pajama: A Mobile Parent and Child Hugging Communication System}},
    year = {2008},
    journal = {International Conference on Interaction Design and Children},
    author = {Keng, James and Teh, Soon and Cheok, Adrian David and Peiris, Roshan L and Choi, Yongsoon and Thuong, Vuong and Lai, Sha},
    pages = {250--257},
    doi = {10.1145/1463689.1463763}
}

@article{Do2021Macro-MiniDisplays,
    title = {{Macro-Mini Actuation of Pneumatic Pouches for Soft Wearable Haptic Displays}},
    year = {2021},
    journal = {IEEE International Conference on Robotics and Automation},
    author = {Do, Brian H. and Okamura, Allison M. and Yamane, Katsu and Blumenschein, Laura H.},
    pages = {14499--14505},
    volume = {},
    publisher = {Institute of Electrical and Electronics Engineers Inc.},
    doi = {10.1109/ICRA48506.2021.9560786}
}

@article{Bartlett2023PeelReview,
    title = {{Peel tests for quantifying adhesion and toughness: A review}},
    year = {2023},
    journal = {Progress in Materials Science},
    author = {Bartlett, Michael D. and Case, Scott W. and Kinloch, Anthony J. and Dillard, David A.},
    pages = {101086},
    volume = {137},
    publisher = {Pergamon},
    doi = {10.1016/J.PMATSCI.2023.101086}
}

@article{Fan2009PilotAmputee,
    title = {{Pilot testing of a haptic feedback rehabilitation system on a lower-limb amputee}},
    year = {2009},
    journal = {ICME International Conference on Complex Medical Engineering},
    author = {Fan, Richard E. and Wottawa, Christopher and Mulgaonkar, Amit and Boryk, Richard J. and Sander, Todd C. and Wyatt, Marilynn P. and Dutson, Erik and Grundfest, Warren S. and Culjat, Martin O.},
    pages = {1--4},
    doi = {10.1109/ICCME.2009.4906637}
}

@article{Kang2023PneumaticSimulation,
    title = {{Pneumatic and acoustic suit: multimodal haptic suit for enhanced virtual reality simulation}},
    year = {2023},
    journal = {Virtual Reality},
    author = {Kang, Daeseok and Lee, Chang Gyu and Kwon, Ohung},
    number = {3},
    pages = {1647--1669},
    volume = {27},
    publisher = {Springer Science and Business Media Deutschland GmbH},
    doi = {10.1007/S10055-023-00756-5}
}

@article{Zhu2020PneuSleeve:Sleeve,
    title = {{PneuSleeve: In-fabric Multimodal Actuation and Sensing in a Soft, Compact, and Expressive Haptic Sleeve}},
    year = {2020},
    journal = {ACM Conference on Human Factors in Computing Systems (CHI)},
    author = {Zhu, Mengjia and Memar, Amirhossein H and Gupta, Aakar and Samad, Majed and Agarwal, Priyanshu and Visell, Yon and Keller, Sean J and Colonnese, Nicholas},
    pages = {206},
    doi = {10.1145/3313831.3376333}
}

@article{Niiyama2015PouchDesign,
    title = {{Pouch motors: Printable soft actuators integrated with computational design}},
    year = {2015},
    journal = {Soft Robotics},
    author = {Niiyama, Ryuma and Sun, Xu and Sung, Cynthia and An, Byoungkwon and Rus, Daniela and Kim, Sangbae},
    number = {2},
    pages = {59--70},
    volume = {2},
    publisher = {Mary Ann Liebert Inc.},
    doi = {10.1089/SORO.2014.0023}
}

@article{Niiyama2014PouchRobotics,
    title = {{Pouch Motors: Printable/inflatable soft actuators for robotics}},
    year = {2014},
    journal = {IEEE International Conference on Robotics and Automation},
    author = {Niiyama, Ryuma and Rus, Daniela and Kim, Sangbae},
    pages = {6332--6337},
    publisher = {Institute of Electrical and Electronics Engineers Inc.},
    doi = {10.1109/ICRA.2014.6907793}
}

@article{Jadhav2023ScalableActuators,
    title = {{Scalable Fluidic Matrix Circuits for Controlling Large Arrays of Individually Addressable Actuators}},
    year = {2023},
    journal = {Advanced Intelligent Systems},
    author = {Jadhav, Saurabh and Glick, Paul E. and Ishida, Michael and Chan, Christian and Adibnazari, Iman and Schulze, Jurgen P. and Gravish, Nick and Tolley, Michael T.},
    number = {},
    pages = {2300011},
    volume = {5},
    publisher = {John Wiley {\&} Sons, Ltd},
    doi = {10.1002/AISY.202300011}
}

@article{Connolly2019Sew-freeRobots,
    title = {{Sew-free anisotropic textile composites for rapid design and manufacturing of soft wearable robots}},
    year = {2019},
    journal = {Extreme Mechanics Letters},
    author = {Connolly, Fionnuala and Wagner, Diana A. and Walsh, Conor J. and Bertoldi, Katia},
    pages = {52--58},
    volume = {27},
    publisher = {Elsevier},
    doi = {10.1016/J.EML.2019.01.007}
}

@article{Rognon2019SoftDrone,
    title = {{Soft Haptic Device to Render the Sensation of Flying Like a Drone}},
    year = {2019},
    journal = {IEEE Robotics and Automation Letters},
    author = {Rognon, Carine and Koehler, Margaret and Duriez, Christian and Floreano, Dario and Okamura, Allison M.},
    number = {3},
    pages = {2524--2531},
    volume = {4},
    publisher = {Institute of Electrical and Electronics Engineers Inc.},
    doi = {10.1109/LRA.2019.2907432}
}

@article{Talhan2024SoftTouch,
    title = {{Soft Pneumatic Haptic Wearable to Create the Illusion of Human Touch}},
    year = {2024},
    journal = {IEEE Transactions on Haptics},
    author = {Talhan, Aishwari and Yoo, Yongjae and Cooperstock, Jeremy R.},
    number = {2},
    pages = {177--190},
    volume = {17},
    publisher = {Institute of Electrical and Electronics Engineers Inc.},
    doi = {10.1109/TOH.2023.3305495}
}

@article{Pohl2017Squeezeback:Notifications,
    title = {{Squeezeback: Pneumatic Compression for Notifications}},
    year = {2017},
    journal = {ACM Conference on Human Factors in Computing Systems (CHI)},
    author = {Pohl, Henning and Brandes, Peter and Quang, Hung Ngo and Rohs, Michael},
    pages = {5318- 5330},
    doi = {10.1145/3025453.3025526}
}

@article{Frediani2020TactileFingertip,
    title = {{Tactile display of softness on fingertip}},
    year = {2020},
    journal = {Scientific Reports},
    author = {Frediani, Gabriele and Carpi, Federico},
    number = {},
    pages = {20491},
    volume = {10},
    publisher = {Nature Publishing Group},
    doi = {10.1038/s41598-020-77591-0}
}

@article{Zhong2006TextilesOverview.,
    title = {{Textiles and human skin, microclimate, cutaneous reactions: an overview.}},
    year = {2006},
    journal = {Cutaneous and Ocular Toxicology},
    author = {Zhong, Wen and Xing, Malcolm M Q and Pan, Ning and Maibach, Howard},
    number = {1},
    pages = {23--39},
    volume = {25},
    doi = {10.1080/15569520500536600}
}

@article{Liu2021ThermoCaress:Stimulation,
    title = {{ThermoCaress: A Wearable Haptic Device with Illusory Moving Thermal Stimulation}},
    year = {2021},
    journal = {ACM Conference on Human Factors in Computing Systems (CHI)},
    author = {Liu, Yuhu and Nishikawa, Satoshi and Ah Seong, Young and Niiyama, Ryuma and Kuniyoshi, Yasuo},
    pages = {214},
    publisher = {ACM},
    doi = {10.1145/3411764.3445777}
}

@article{Wu2010Torso-basedDisorders,
    title = {{Torso-based tactile feedback system for patients with balance disorders}},
    year = {2010},
    journal = {IEEE Haptics Symposium},
    author = {Wu, Steven W. and Fan, Richard E. and Wottawa, Christopher R. and Fowler, Eileen G. and Bisley, James W. and Grundfest, Warren S. and Culjat, Martin O.},
    pages = {359--362},
    doi = {10.1109/HAPTIC.2010.5444630}
}

@article{Nolan1982Two-pointWomen,
    title = {{Two-point discrimination assessment in the upper limb in young adult men and women}},
    year = {1982},
    journal = {Physical Therapy},
    author = {Nolan, M. F.},
    number = {7},
    pages = {965--969},
    volume = {62},
    publisher = {Phys Ther},
    doi = {10.1093/PTJ/62.7.965}
}

@article{Ercalk2021Two-pointIndividuals,
    title = {{Two-point discrimination assessment of the lower extremities of healthy young Turkish individuals}},
    year = {2021},
    journal = {Somatosensory {\&} Motor Research},
    author = {Er{\c{c}}alık, Cem and {\"{O}}zkurt, Seçil},
    number = {3},
    pages = {253--257},
    volume = {38},
    publisher = {Somatosens Mot Res},
    doi = {10.1080/08990220.2021.1959310}
}

@article{VanBeek2024ValidationFeedback,
    title = {{Validation of a Soft Pneumatic Unit Cell (PUC) in a VR Experience: A Comparison between Vibrotactile and Soft Pneumatic Haptic Feedback}},
    year = {2024},
    journal = {IEEE Transactions on Haptics},
    author = {Van Beek, Femke E. and Bisschop, Quinten P.I. and Kuling, Irene A.},
    number = {2},
    pages = {191--201},
    volume = {17},
    publisher = {Institute of Electrical and Electronics Engineers Inc.},
    doi = {10.1109/TOH.2023.3307872}
}

@article{Wu2019WearableArm,
    title = {{Wearable Haptic Pneumatic Device for Creating the Illusion of Lateral Motion on the Arm}},
    year = {2019},
    journal = {IEEE World Haptics Conference},
    author = {Wu, Weicheng and Culbertson, Heather},
    pages = {193--198},
    publisher = {Institute of Electrical and Electronics Engineers Inc.},
    doi = {10.1109/WHC.2019.8816170}
}

@article{Raitor2017WRAP:Guidance,
    title = {{WRAP: Wearable, restricted-aperture pneumatics for haptic guidance}},
    year = {2017},
    journal = {IEEE International Conference on Robotics and Automation},
    author = {Raitor, Michael and Walker, Julie M. and Okamura, Allison M. and Culbertson, Heather},
    pages = {427--432},
    publisher = {Institute of Electrical and Electronics Engineers Inc.},
    doi = {10.1109/ICRA.2017.7989055}
}
\end{document}